\documentclass[11pt,a4paper,paper]{JHEP3}
\usepackage{amsmath}
\usepackage{amssymb}
\usepackage{colordvi}
\usepackage{axodraw}
\usepackage{epsfig}
\usepackage{subfigure}
\newcommand{\nn}{\nonumber}
\newcommand{\Frac}[2]{\frac{\displaystyle{#1}}{\displaystyle{#2}}}
\newcommand{\lsim}{\raise0.3ex\hbox{$\;<$\kern-0.75em\raise-1.1ex\hbox{$\sim\;$}}}
\newcommand{\dir}[1]{\; /\hspace{-0.7em}#1}
\newcommand{\gsim}{\raise0.3ex\hbox{$\;>$\kern-0.75em\raise-1.1ex\hbox{$\sim\;$}}}
\newcommand{\eq}[1]{Eq.~(\ref{#1})}
\newcommand{\fig}[1]{Fig.\ \ref{#1}}
\newcommand{\unity}{{\hbox{1\kern-.8mm l}}}
\newcommand{\abs}[1]{\left|#1\right|}

\newcommand{\WRu}{{\mathbf{W_R^u}}}
\newcommand{\WRd}{{\mathbf{W_R^d}}}
\newcommand{\WL}{{\mathbf{W_L}}}
\newcommand{\WRudag}{{\mathbf{W_R^u}}^{\dagger}}
\newcommand{\WRddag}{{\mathbf{W_R^d}}^{\dagger}}
\newcommand{\WLdag}{{\mathbf{W_L}}^{\dagger}}
\newcommand{\Yu}{{Y_u^{\phantom{\dagger}}}}
\newcommand{\Yud}{{Y_u^{\dagger}}}
\newcommand{\Yd}{{Y_d^{\phantom{\dagger}}}}
\newcommand{\Ydd}{{Y_d^{\dagger}}}
\newcommand{\Yx}{{Y_x^{\phantom{\dagger}}}}
\newcommand{\Yxd}{{Y_x^{\dagger}}}
\newcommand{\UxL}{{\mathcal U^x_L}}
\newcommand{\UuL}{{\mathcal U^u_L}}
\newcommand{\UdL}{{\mathcal U^d_L}}
\newcommand{\UxR}{{\mathcal U^x_R}}
\newcommand{\UuR}{{\mathcal U^u_R}}
\newcommand{\UdR}{{\mathcal U^d_R}}
\newcommand{\UQ}{{\mathcal U^Q}}
\newcommand{\UxLdag}{{\mathcal U^x_L}^{\dagger}}
\newcommand{\UuLdag}{{\mathcal U^u_L}^{\dagger}}
\newcommand{\UdLdag}{{\mathcal U^d_L}^{\dagger}}
\newcommand{\UxRdag}{{\mathcal U^x_R}^{\dagger}}

\newcommand{\UQdag}{{\mathcal U^Q}^{\dagger}}
\newcommand{\V}[1]{V_{#1}^{\phantom{\ast}}}
\newcommand{\Vc}[1]{V_{#1}^{\ast}}
\newcommand{\U}[1]{U_{#1}^{\phantom{\ast}}}
\newcommand{\Uc}[1]{U_{#1}^{\ast}}
\newcommand{\PxL}[1]{{\mathsf P_{#1}^{x_L}}}
\newcommand{\PxR}[1]{{\mathsf P_{#1}^{x_R}}}
\newcommand{\PuL}[1]{{\mathsf P_{#1}^{u_L}}}

\newcommand{\PdL}[1]{{\mathsf P_{#1}^{d_L}}}
\newcommand{\PdR}[1]{{\mathsf P_{#1}^{d_R}}}
\newcommand{\PQ}[1]{{\mathsf P_{#1}^{Q}}}
\newcommand{\Pcan}[1]{\mathbf{P_{#1}}}
\newcommand{\Jota}[3]{{J^{(#1)}_{#2,#3}}}
\newcommand{\IInv}[2]{{I_{#1,#2}}}
\newcommand{\masaX}[1]{{m_{x_#1}^2}}
\newcommand{\masaU}[1]{{m_{u_#1}^2}}
\newcommand{\masaD}[1]{{m_{d_#1}^2}}
\newcommand{\masaQ}[1]{{m_{\tilde Q_#1}^2}}
\newcommand{\HQ}{{H_{\tilde Q}}}
\newcommand{\tr}[1]{{\text{Tr}\left(#1\right)}}
\newcommand{\im}[1]{{\text{Im}\left[#1\right]}}
\newcommand{\re}[1]{{\text{Re}\left[#1\right]}}
\newcommand{\imtr}[1]{{\im{\tr{#1}}}}
\newcommand{\KET}[2]{\left|#1_{L#2}\right>}
\newcommand{\BRA}[2]{\left<#1_{L#2}\right|}

\newcommand{\citeflavour}{\cite{flavour}}
\newcommand{\citeMSSMCP}{\cite{MSSMCP}}
\newcommand{\citeSUSYCollider}{\cite{SUSYCollider}}
\newcommand{\citeINVARIANTS}{\cite{Botella:1995cs,Branco:1989kf,INVARIANTS,Botella:1986gb,Jarlskog:1987zd,Branco:1999fs,Lebedev:2002wq}}
\newcommand{\citeCKM}{\cite{CKM}}
\newcommand{\citeSUSYReview}{\cite{SUSYReview}}
\newcommand{\citecMSSM}{\cite{cMSSM,Bartl:2001wc}}
\newcommand{\citeMFV}{\cite{MFV}}
\newcommand{\citeSUSYFlavour}{\cite{SUSYFlavour}}
\newcommand{\citeCKMfits}{\cite{CKMfits}}
\newcommand{\citeSUSYKMix}{\cite{Masiero:2000ni,Akama:2001em}}
\newcommand{\citeSUSYBPsiK}{\cite{Masiero:2000ni,SUSYBPsiK}}
\newcommand{\citeSUSYBsMix}{\cite{SUSYBsMix}}
\newcommand{\citeBsgExp}{\cite{BsgExp}}
\newcommand{\citeBsgTeo}{\cite{BsgTeo,Bartl:2001wc}}
\newcommand{\citeKPiNuNu}{\cite{KPiNuNu}}
\newcommand{\citeMI}{\cite{MI}}

\title{Invariant approach to flavour-dependent CP-violating phases in the MSSM}
\author{F.J. Botella, M. Nebot\\ Departament de F\'{\i}sica Te\`orica and IFIC, Universitat de Val\`encia-CSIC, E-46100, Burjassot, Spain\\ \email{Francisco.J.Botella@uv.es, Miguel.Nebot@uv.es}}
\author{O. Vives\\ Department of Physics, TH Division, CERN\\
Geneva 23, Switzerland\\ \email{Oscar.Vives@cern.ch}}

\abstract{We use a new weak basis invariant approach to classify all the
observable phases in any extension of the Standard Model (SM). We
apply this formalism to determine the invariant CP phases in a
simplified version of the Minimal Supersymmetric SM with only three
non-trivial flavour structures. We propose four experimental measures
to fix completely all the observable phases in the model.  After 
these phases have been determined from experiment, we are able to make 
predictions on any other CP-violating observable in the theory, much in 
the same way as in the Standard Model all CP-violation observables are 
proportional to the Jarlskog invariant.}

\preprint{hep-ph/0407349}
\preprint{CERN-PH-TH/2004-138\\ IFIC/04-41\\ FTUV-04-0730}

\begin{document}

%\vspace*{-3cm}
%\begin{flushright}
%hep-ph/0407349 \\
%CERN-PH-TH/2004-138\\
%IFIC/04-41\\
%FTUV-04-0730\\
%July 2004
%\end{flushright}
                                                                                
%\begin{center}
%\begin{Large}
%{\bf Invariant approach to flavour-dependent\\ CP-violating phases in the MSSM}
%\end{Large}
                                                                                
%\vspace{0.5cm}
%F.J. Botella $^a$, M. Nebot $^a$ and O. Vives $^b$\\[0.2cm]
%{\it $^a$ Departament de F\'{\i}sica Te\`orica and IFIC\\
%Universitat de Val\`encia-CSIC, E-46100, Burjassot, Spain}\\
%{\it $^b$ Department of Physics, TH Division, CERN\\
%Geneva 23, Switzerland}
%\end{center}

%\begin{abstract}
%We use a new weak basis invariant approach to classify all the
%observable phases in any extension of the Standard Model (SM). We
%apply this formalism to determine the invariant CP phases in a
%simplified version of the Minimal Supersymmetric SM with only three
%non-trivial flavour structures. We propose four experimental measures
%to fix completely all the observable phases in the model.  After 
%these phases have been determined from experiment, we are able to make 
%predictions on any other CP-violating observable in the theory, much in 
%the same way as in the Standard Model all CP-violation observables are 
%proportional to the Jarlskog invariant.
%\end{abstract}
%%%%%%%%%%%%%%%%%%%%%%%%%%%%%%%%%%%%%%%%%%%%%%%%%%
\section{Introduction}\label{SEC:INTRO}
%%%%%%%%%%%%%%%%%%%%%%%%%%%%%%%%%%%%%%%%%%%%%%%%%%

From the point of view of theory, the origin of flavour and CP
violation constitute two of the most urging questions still unanswered
in high energy physics. Different theoretical ideas have been proposed
to improve our understanding of these problems 
\citeflavour
, but new
experimental input is urgently required to unravel this complex
puzzle. In the interlude between the LEP and LHC colliders, CP violation
and Flavour-Changing-Neutral-Current (FCNC) experiments at low
energies are now the main field of research. Even after the start of
the LHC, the interplay between the information obtained at the LHC and the
information from indirect searches will play a fundamental role in the
understanding of CP violation and flavour.  

In the Standard Model (SM) both problems are deeply related and the
only source of both CP violation and flavour lies in the fermionic
Yukawa couplings. In a three-generation SM there is only a single CP-odd quantity invariant under redefinitions of the quark basis.  This
CP-violating quantity has a nice weak basis invariant formulation with
the well-known Jarlskog invariant \cite{Jarlskog:1985ht,Jarlskog:1985cw,Bernabeu:1986fc,Roldan:1991,Botella:1995cs}:
\begin{multline}
J_{\text{CP}}=
\det\left(-i\left[\Yu \Yud,\Yd \Ydd\right]\right)=\frac{i}{3}\tr{\left[\Yu \Yud,\Yd \Ydd\right]^3}=\\
-2~\imtr{(\Yu \Yud) (\Yd \Ydd) (\Yu \Yud)^2(\Yd \Ydd)^2}
\end{multline}
and all CP-violation effects in the SM are associated to this single
observable phase.
In general, any extension of the SM includes additional sources of CP
violation and new flavour structures, which increase the number of
observable phases. Supersymmetry is perhaps the most complete and
(theory-) motivated extension of the Standard Model, and we expect to
be able to find the supersymmetric particles in the neighbourhood of
the electroweak scale. The Minimal Supersymmetric Standard Model
(MSSM) is a perfect example of the increase in the number of
observable phases in extensions of the SM.  The number of parameters
in a generic MSSM, including real flavour parameters and CP phases, is
of 124 \cite{Haber:1998if}, out of which there are 44 physical phases.  Most
of these phases have received no attention until recently and only two
of them, namely the relative phase between the gaugino masses
and the global trilinear phase $\varphi_A = \mbox{Arg}(A^*M)$ with
$\mbox{Arg}(B \mu)=0$ and the relative phase between the $\mu$-term in
the superpotential and the gaugino masses, $\varphi_\mu =
\mbox{Arg}(\mu^*M)$, have received full attention, thanks to their
relation with electric dipole moments (EDMs). However, a generic MSSM
introduces many additional mixings and phases, and these parameters
have important effects in FCNC and CP-violation experiments 
\citeMSSMCP. In collider experiments, the presence of SUSY
phases can also have a measurable impact 
\citeSUSYCollider. Therefore, if SUSY
is found either directly or indirectly in near-future experiments, all
the SUSY phases will become observable and a classification of these
phases and the construction of fermion basis invariants analogous to the
Jarlskog invariant become especially important\footnote{Weak bases invariants in the context of the MSSM were originally used by Branco and Kostelecky to find the necessary and sufficient conditions for CP conservation in \cite{Branco:1989kf}.}.

In this work we develop a complete formalism that generalizes the
construction of weak basis invariants to a generic extension of the
SM. The construction of invariants under weak basis transformations
(WBTs) to study CP violation in the SM and its extensions has been
undertaken for a long time 
\citeINVARIANTS. In this work we
extend these analyses by introducing an improved formalism to relate
these invariants with observables directly measurable at future
experiments. This formalism allows us to translate directly the usual
Feynman diagrams into weak basis invariants and
vice versa. Furthermore, we are able to define a basis of independent
weak basis invariants and to express any invariant in terms of this
basis. This program was partly developed in \cite{Botella:1995cs} and
\cite{Lebedev:2002wq} without the necessary connexion to experimental
observables. In particular, in \cite{Lebedev:2002wq} a set of weak basis
invariants spanning all the observable phases in the quark sector of a
general MSSM was constructed, but the connexion to experimental
observables and relations between them was not explicitly presented.

The outline of this work is as follows. In section \ref{SEC:wbt:ri}
we define weak basis transformations and present our formalism to
build weak basis invariants. We apply this formalism to several simple
examples in the SM. In section \ref{SEC:MSSM} we analyse the quark
sector of a simple MSSM with only three flavour structures, the Yukawa
matrices and the squark doublet mass matrix taking all other matrices to be
universal. In this simple model we show the full power of our
formalism. We find a basis of independent invariants and select a set
of experimental observables to fix these invariants. Finally we show
that any other CP-violating observable in this model is completely
fixed in terms of our basis of invariants, masses and moduli of mixing
angles. Therefore, we are able to make predictions on any other CP-violating observable in the theory.

%%%%%%%%%%%%%%%%%%%%%%%%%%%%%%%%%%%%%%%%%%%%%%%%%%
\section{Weak basis transformations and\\ rephasing invariance in the SM}
\label{SEC:wbt:ri}
%%%%%%%%%%%%%%%%%%%%%%%%%%%%%%%%%%%%%%%%%%%%%%%%%%

The SM lagrangian density $\mathcal L_{\text{SM}}$ includes the following
$SU(3) \otimes SU(2)_L\otimes U(1)_Y$ contributions
\begin{align}
\mathcal L_{G+F+H}&=-\frac{1}{4} G_{\mu\nu}^b G^{\mu\nu}_b
-\frac{1}{4} W_{\mu\nu}^a
W^{\mu\nu}_a-\frac{1}{4}B_{\mu\nu}B^{\mu\nu}+\bar q_L^0 i\dir D
q_L^0\nn \\ &+\bar u_R^0 i\dir D u_R^0+\bar d_R^0 i\dir D d_R^0+(D_\mu
\Phi)^\dagger (D^\mu \Phi)-V(\Phi) \\ \mathcal L_{H+F}&=-\bar q_L^0
Y_u u_R^0\tilde\Phi-\bar q_L^0 Y_d d_R^0 \Phi+\text{h.c.},
\label{lag:SM}
\end{align}
with $D_\mu$ the covariant derivative and the fields
$q_L^0,u_R^0,d_R^0$ in an arbitrary basis with non-diagonal Yukawa
couplings. In the SM, we have three copies (i.e. generations) of
representations of the $SU(3)\otimes SU(2)_L\otimes U(1)_Y$ gauge
group with different masses and mixing angles given by $\mathcal
L_{H+F}$. However $\mathcal L_{G+F+H}$ only depends on the gauge
quantum numbers, and it is completely independent of how we label these
three copies; it is invariant under $U(3)_L\otimes U(3)_{u_R}\otimes
U(3)_{d_R}$ global transformations acting on $q_L,~u_R$ and
$d_R$. These global transformations are WBTs. Two equivalent field assignments are related by
\begin{equation}
 q^0_L=\WL {q^0_L}^\prime \quad ; \quad u^0_R=\WRu {u^0_R}^\prime
 \quad ; \quad d^0_R=\WRd {d^0_R}^\prime~, \label{WBT:01}
\end{equation}
where $\WL \in U(3)_L$, $\WRu \in U(3)_{u_R}$ and $\WRd \in
U(3)_{d_R}$. While $\mathcal L_{G+F+H}$ is explicitly invariant under
these $U(3)_L\otimes U(3)_{u_R}\otimes U(3)_{d_R}$ WBTs, $\mathcal
L_{H+F}$ is not invariant. Under WBTs, $\mathcal L_{H+F}$ changes to
\begin{equation}
\mathcal L^\prime_{H+F}= - \bar q_L^{0\prime}~\Yu ~
{u^{0\prime}_R} \tilde\Phi-\bar q_L^{0\prime}~\Yd
~{d^{0\prime}_R}\Phi+\text{h.c.}~,\label{WBT:01b}
\end{equation}
where the fields transform as in \eq{WBT:01} and the Yukawa couplings
are unchanged. However, if we allow these Yukawa couplings to
transform under WBTs as
\begin{equation}
\Yu \to \Yu^\prime=\WLdag \Yu \WRu \quad ; \quad \Yd \to \Yd^\prime=\WLdag \Yd \WRd~,\label{WBT:02}
\end{equation}
then $\mathcal L_{H+F}=\mathcal L^\prime_{H+F}$ and the full $\mathcal
L_{\text{SM}}$ is invariant under WBTs. Therefore the two theories given by
$\{{q^0_L}^\prime,{u^0_R}^\prime,{d^0_R}^\prime,Y_u^\prime,Y_d^\prime\}$
and $\{q^0_L,u^0_R,d^0_R,Y_u,Y_d\}$ provide equivalent
\emph{physics}. These two theories correspond to two different choices
of the weak basis that we use to formulate our theory. Clearly any
physical observable should be independent of our choice and thus weak
basis invariant. Physical processes involve quarks with definite mass. The mass eigenstates basis is defined by diagonalizing the Yukawa couplings through biunitary transformations\footnote{$v$ is the spontaneous symmetry breaking VEV of $\Phi$.}
\begin{equation}
\UuLdag \Yu \UuR=\frac{1}{v}M_u \quad ; \quad \UdLdag \Yd \UdR=\frac{1}{v}M_d ~.\label{DIAG:01}
\end{equation}
They correspond to the change of basis
\begin{equation}
u^0_L=\UuL u_L \qquad d^0_L=\UdL d_L \qquad u^0_R=\UuR u_R \qquad d^0_R=\UdR d_R ~,\label{DIAG:02}
\end{equation}
where $u_L,u_R,d_L$ and $d_R$ are the mass eigenstates. Under this diagonalization the Cabibbo--Kobayashi--Maskawa matrix 
\citeCKM, $\UuLdag \UdL\equiv V$ appears in the charged-current couplings. Going to the mass basis in the SM corresponds to the following reparametrization of the lagrangian:
\[
(\Yu,\Yd)\to \left(V,\frac{1}{v}M_u,\frac{1}{v}M_d\right).
\]
It must be stressed that the transformation \eq{DIAG:02} \emph{is not} a WBT: $u_L$ and $d_L$ transform independently. Nevertheless this reparametrization, as is well known, is remarkably useful, among other facts because this new set of parameters is WBT-invariant. First we notice that the biunitary transformations diagonalizing the Yukawa matrices also change under a WBT:
\begin{equation}
\UuL^\prime=\WLdag ~\UuL \qquad \UdL^\prime=\WLdag ~\UdL \qquad \UuR^\prime=\WRudag ~\UuR \qquad \UdR^\prime=\WRddag ~\UdR \label{WBT:03}
\end{equation}
Then, we have
\begin{align}
V^\prime &=\UuL^{\prime\dagger}~\UdL^\prime=\UuLdag ~\WL \WLdag ~\UdL=\UuLdag~\UdL =V\nn\\
\frac{M_u^\prime}{v} &=\UuL^{\prime\dagger}~ \Yu^\prime ~\UuR^\prime=\UuLdag ~\WL \WLdag~ \Yu ~\WRu\WRudag~ \UuR=~\UuLdag~ \Yu~ \UuR = \frac{M_u}{v}\nn\\
\frac{M_d^\prime}{v} &=\UdL^{\prime\dagger}~ \Yd^\prime ~\UdR^\prime=\UdLdag ~\WL \WLdag~ \Yd ~\WRd\WRddag~ \UdR=~\UdLdag~ \Yd~ \UdR = \frac{M_d}{v}\label{WBT:04}
\end{align}
and thus they are clearly weak basis invariants. Nevertheless this common parametrization is not fully defined by \eq{DIAG:02}: as in any diagonalization, the phases of the mass eigenstates are not well defined. This freedom can be incorporated to \eq{DIAG:02} with the following generalization
\begin{equation}
u^0_L=\UuL e^{i\Theta^u_L} u_L \qquad d^0_L=\UdL e^{i\Theta^d_L} d_L \qquad u^0_R=\UuR e^{i\Theta^u_R} u_R \qquad d^0_R=\UdR e^{i\Theta^d_R} d_R \label{DIAG:03}
\end{equation}
with $\Theta^u_L,\Theta^u_R,\Theta^d_L,\Theta^d_R$ real, diagonal matrices. It is precisely this freedom that allows choosing the masses real and positive. Even after this choice, we still have some rephasing freedom that explicitly keeps the diagonal elements in the mass matrix real and positive. This corresponds to rephasing the mass eigenstates with 
$\Theta^u_L=\Theta^u_R\equiv\Theta^u$, $\Theta^d_L=\Theta^d_R\equiv
\Theta^d$; mass eigenvalues are then invariant under those (reduced)
rephasings that we consider in the rest of this work. Under the above-mentioned rephasing transformations, $V$ is not invariant and it goes
to $e^{-i\Theta^u}~V~e^{i\Theta^d}$, implying $\V{jk}\to
\V{jk}e^{i(\theta^d_k-\theta^u_j)}$. These are the famous rephasings of
the CKM matrix that reduce to a single phase the number of physical
phases in a three-generation SM. As a by-product we conclude that physical observables must be both WBT-invariant and rephasing-invariant.
Notice, for example, that $V$ is WBT-invariant and rephasing-\emph{variant}, and thus $V$ will necessarily enter observables through rephasing-invariant combinations, as in \eq{INV:00}. 

In models where all the flavour couplings in the lagrangian are
bilinear in the flavoured fields (as in the SM), WBT and rephasing
invariance automatically imply that any physical observable can be
written in terms of traces of well-behaved products of flavour
matrices. If we define
\begin{equation}
H_u\equiv v^2\Yu\Yud \quad,\quad H_d\equiv v^2\Yd\Ydd\quad,\quad H_i\overset{\text{\tiny WBT}}{\to} H_i^\prime =\WLdag~H_i~\WL~,\label{WBT:05} 
\end{equation}
it is well known that any physical observable can be written in terms of:
\begin{equation}
\tr{(H_u)^a(H_d)^b(H_u)^c(H_d)^d\ldots}. \label{INV:00}
\end{equation}
We will call these structures weak basis invariants (WBI).
Note that the only matrix transforming under $U(3)_{u_R}$ or
$U(3)_{d_R}$ is $Y_j^\dagger Y_j^{\phantom{\dagger}}$ ($j=u,d$) and
therefore, with this matrix, we can only construct the trivial
observable
\[
\tr{(Y_j^\dagger Y_j^{\phantom{\dagger}})^a}=\frac{1}{v^{2a}}\tr{(M_j)^{2a}}=\frac{1}{v^{2a}}\sum_{k}(m_{j_k})^{2a},
\]
implying that right-handed rotations are not observable. 

For CP violation it is clear that
$\imtr{(H_u)^a(H_d)^b(H_u)^c(H_d)^d\ldots}$ is a genuine CP-violating
phase. Obtaining the Jarlskog invariant \cite{Jarlskog:1985ht} in the SM is an
instructive exercise. As $H_j$ is hermitian,
$\imtr{H_j}=\imtr{H_jH_k}=0$; the first invariant with an imaginary
part different from zero is
\begin{multline}
J=\imtr{H_u H_d H_u^2 H_d^2}=(m_t^2-m_c^2)(m_t^2-m_u^2)(m_c^2-m_u^2)\\
\times (m_b^2-m_s^2)(m_b^2-m_d^2)(m_s^2-m_d^2)~ \im{\V{22}\Vc{23}\V{33}\Vc{32}}. \label{INV:J}
\end{multline}
Invariants like \eq{INV:00} and its generalization are quite useful to
find out the necessary and sufficient conditions to have CP violation
in a given model and therefore to find out the number of independent
CP-violating phases \cite{Branco:1999fs}. Nevertheless its relation with physical
observables is far from obvious. In the SM, $J$ only appears in
observables ``averaged'' over all the quarks, as in the case of the CKM contribution to the 
electric dipole moments (EDMs) of leptons; \eq{INV:J} never appears in
its full glory in CP-violating observables of the quark sector. The
reason why $J$ does not appear in CP-violating observables of the
quark sector is clear. Equation (\ref{INV:J}) encodes all the necessary conditions
to have CP violation, but, if we are able to
distinguish a $b$ quark from an $s$ quark experimentally, then a given CP-violating
observable involving both quarks does not require the presence of the
factor $(m_b^2-m_s^2)$. Therefore there must be a way, much simpler than
\eq{INV:J}, of writing WBIs directly related to physical observables.

The key point to reach this goal is to write $H_j$ in terms of
projection operators over the mass eigenstates, i.e.
\begin{equation}
H_u=\sum_{i=1}^3\masaU{i}\KET{u}{i}\BRA{u}{i}=\sum_{i=1}^3\masaU{i}\PuL{i}.\label{H:decomp}
\end{equation}
From \eq{DIAG:02} it is evident that $\left[\PuL{i}\right]_{\alpha\beta}=\left(\UuL\right)_{\alpha i}\left(\UuLdag\right)_{i\beta}$, that is
\begin{equation}
\PuL{i}=\UuL \Pcan{i} \UuLdag\quad ; \quad \left(\Pcan{i}\right)_{jk}=\delta_{ij}\delta_{ik}~.\label{Proj:01}
\end{equation}
These projection operators transform under WBTs (see \eq{WBT:03}) as $H_u$. It is worthwhile to mention that, given $\Yu$, $H_u$ is perfectly defined and so are $\UuL$ and $\PuL{i}$. In general we can define the following chiral projectors with well-defined WBT properties:
\begin{equation}
\PxL{i}=\UxL \Pcan{i} \UxLdag \quad ; \quad \PxL{i}\to \PxL{i}^\prime=\WLdag ~ \PxL{i} ~ \WL~,\qquad x=u,d. \label{Proj:02}
\end{equation}
The WBIs in \eq{INV:00} are generalized by allowing any substitution $H_u\to\PuL{i}$ and $H_d\to\PdL{i}$. For right-handed projectors $\PxR{i}=\UxR \Pcan{i} \UxRdag$, $x=u,d$, the following relation 
\begin{equation}
v^2 \Yx\PxR{i}\Yxd=(\UxL M_x \UxRdag)(\UxR \Pcan{i} \UxRdag)(\UxR M_x \UxLdag)=\masaX{i} \PxL{i}\qquad x=u,d \label{RightProj}
\end{equation}
reflects the inobservability of right-handed rotations and reproduces
the well known result that the only
thing we need to introduce a right-handed field is a mass insertion. Note that once we use a
right-handed projector, it is mandatory to have, inside \eq{RightProj}, the string $\Yx\PxR{i}\Yxd$. Equation (\ref{RightProj}) allows us to avoid right-handed projectors.

By using projection operators, the most simple WBI we can construct is
\begin{equation}
\tr{\PuL{i}\PdL{j}}=\abs{\V{ij}}^2\propto \Gamma(d_{Lj}\to u_{Li}~W),\label{OBS:00}
\end{equation}
where the first equality is obtained using \eq{Proj:01} and the last
proportionality is trivial from the previous result, but can also be
obtained from the presence of two projectors, which means the square of
the amplitude $\BRA{u}{i}\mathbf{1}\KET{d}{j}$. Note that in the weak
basis where we are working, the flavour structure of the $W$ coupling is just
the identity. This example shows that, using projection operators, one
can write much simpler WBIs, directly related to
physical processes. The first kind of CP-violating WBI made
with projectors is
\begin{multline}
\imtr{\PuL{1}\PdL{1}\PuL{2}\PdL{2}}=\im{\V{11}\Vc{21}\V{22}\Vc{12}}\propto\\
\propto\Gamma\left(D_s^+\to K^0\pi^+\right)-\Gamma\left(D_s^-\to\bar K^0\pi^-\right).\label{OBS:01}
\end{multline}

%\begin{multline}
%\Gamma\left(D_s^+\to K^0\pi^+\right)-\Gamma\left(D_s^-\to\bar K^0\pi^-\right)\propto\imtr{\PuL{1}\PdL{1}\PuL{2}\PdL{2}}\\ =\im{\V{11}\Vc{21}\V{22}\Vc{12}}. \label{OBS:01}
%\end{multline}
This WBI is the well celebrated imaginary part of the
quartets of the CKM matrix. It must be related to the CP-violating
interference of two different weak amplitudes that appear at tree
level, because there are no internal masses. This is the case for
$D_s^+\to K^0\pi^+$, where the interfering amplitudes are the decay
$c\to ud\bar d$ and the annihilation $(c\bar s\to u\bar s)^\ast$; it
is clear that in this interference we have exactly the four projectors
in \eq{OBS:01}: $u\bar u$, $d\bar d$, $c\bar c$ and $s\bar s$. It is
worthwhile to mention that, from the experimental side, in this decay
we are tagging the quarks $u$, $d$, $c$ and $s$, and there is therefore no
need for any mass suppression factor such as the ones in
\eq{INV:J}~\footnote{Note that in the decay $D_s^+\to K^0\pi^+$ there
is also a highly suppressed penguin contribution.}. Notice that, in
addition, non-zero strong phase differences are required. WBIs especially useful are those that involve projectors and the
flavour structures of the lagrangian as $H_u$ and $H_d$. These invariants 
are discussed in the next section in a wider context.

%%%%%%%%%%%%%%%%%%%%%%%%%%%%%%%%%%%%%%%%%%%%%%%%%%
\section{The Minimal Supersymmetric SM}\label{SEC:MSSM}
%%%%%%%%%%%%%%%%%%%%%%%%%%%%%%%%%%%%%%%%%%%%%%%%%%
In this section we analyse WBIs in the MSSM. If no
interactions connect leptons and quarks, as is the case of an R-parity-conserving MSSM, we can consider both sectors separately.  In this
work we concentrate on quarks. A general MSSM involves 7 independent
flavour matrices in the quark sector 
\citeSUSYReview. These 7 flavour matrices are:
\begin{equation}
\Yu,\qquad \Yd,\qquad Y_u^A,\qquad Y_d^A,\qquad M_Q^2,\qquad M_U^2,\qquad M_D^2. \label{MSSM:Fl:Mat:00}
\end{equation}
In some scenarios, such as the so-called Constrained MSSM \citecMSSM~or Minimal
Flavour Violation models, the soft mass matrices are supposed to be
universal and the trilinear matrices proportional to the Yukawa
matrices 
\citeMFV. However, in realistic supersymmetric flavour models 
\citeSUSYFlavour~we
expect all these matrices to have non-trivial flavour structures.
Here, as a first approach to this enlarged flavour scenario, we will
consider a simplified situation with 3 non-trivial flavour matrices; in
this restricted MSSM, $\Yu$, $\Yd$, and $M_Q^2$ are generic matrices,
while the remaining matrices formula (\ref{MSSM:Fl:Mat:00}) are:
\begin{eqnarray}
& M_U^2=m_{\tilde u}^2 \mathbf{1}\qquad & M_D^2=m_{\tilde d}^2
\mathbf{1} \nn \\ & Y_u^A=A_0\Yu=A_0^\ast\Yu\qquad &
Y_d^A=A_0\Yd=A_0^\ast\Yd \label{MSSM:Fl:Mat:01}
\end{eqnarray}
where $m_{\tilde u}^2$, $m_{\tilde d}^2$ and $A_0$ are real numbers;
$\HQ\equiv M_Q^2$ is hermitian and, under a WBT, transforms as $H_u$ and
$H_d$ in \eq{WBT:05}. Furthermore, as we are mainly interested in
flavour-dependent phases we also take a real $\mu$ parameter in the
superpotential. Using the same strategy as was already used for the SM in the
previous section, we can build a complete set of invariants in our
simple MSSM model. The number of independent parameters, and thus of
independent observables, can be determined as shown in \cite{Santamaria:1993ah}:
\begin{equation}
N=N_{Fl}-N_G+N_{G^\prime},
\label{Parameter:count}
\end{equation}
where $N_{Fl}$ is the number of parameters in the flavour matrices,
$N_G$ is the number of parameters of the WBTs group $G=U(3)_L\otimes
U(3)_{u_R}\otimes U(3)_{d_R}$, and $G^\prime$ is the subgroup of $G$
under which the flavour matrices are invariant, that is the subgroup
of $G$ unbroken by the flavour matrices. Equation (\ref{Parameter:count}) applies
separately to mixings+masses and to phases. In our simple MSSM, we
have $N_{Fl}=2 \times 18 + 9 = 45$, $N_G = 3\times 9 = 27$ and the
unbroken subgroup is only $U(1)$ corresponding to baryon number
conservation, $N_{G^\prime}=1$.  Therefore this yields 9 masses, $3\times 2$ mixing
angles and 4 CP-violating phases. Corresponding to the 4 CP-violating
phases in the model we only need 4 independent complex invariants to
describe CP violation. Using the formalism developed in the previous
section, it is clear that we can build an infinite number of complex
invariants. However, as we prove in Appendix \ref{APP:traces}, we can
always express any invariant in terms of a chosen set of four
independent invariants.

A second ingredient needed to determine all the independent phases in our model is the possibility to relate these independent invariants to physical observables. As we saw in the previous section, this is achieved through the introduction of projectors on external states. In fact, we can make a direct correspondence between these invariants and Feynman-like flavour diagrams. Starting from a (set of) Feynman diagram(s) we can immediately read the corresponding WBI(s) and, conversely, we can draw the Feynman diagrams corresponding to a given invariant. To do this we only need a few considerations:
\begin{itemize}
\item Full invariants and physical observables always correspond to cross sections or decay rates and hence squared moduli of amplitudes (Feynman diagrams). 
\item We obtain a flavour loop by joining a Feynman diagram with a conjugated Feynman diagram contributing to the same amplitude. In theories where the couplings are always bilinear in flavour we obtain a closed {\em flavour path} corresponding directly to a trace. 
\item Each initial or final particle is represented by a projector on the corresponding mass eigenstate.
\item To every virtual particle in loops we associate a full flavour structure, $H_u$, $H_d$ or $H_{\tilde Q}$. Strictly speaking we must include an arbitrary function of those matrices, which can be expanded: $f(H_{X},H_Y\ldots)=\sum_{nm}C_{nm}H_{X}^n H_Y^m\dots$.
\item In a given \emph{flavour path} any transition between different flavour matrices (projectors or full flavour structures) is mediated by the appropriate flavour-blind gauge or gaugino lines taking into account the charge and spin of the particles involved: $W^+ \equiv H_{u} \leftrightarrow H_d$,~$\chi^0,~\tilde g\equiv  H_{u} \leftrightarrow H_{\tilde Q}(u)$,~~$\chi^+\equiv  H_{u} \leftrightarrow H_{\tilde Q}(d)$, \dots
\item Neutral gauge bosons do not modify flavour, hence an arbitrary number of them may be attached at any point of our \emph{flavour path}.
\end{itemize}

Using these as rules, it is straightforward to translate invariants into
Feynman diagrams and vice versa. However, it is important to notice
that a given invariant may correspond to different processes. For
instance adding any number of external photons, gluons or $Z$ bosons to
our Feynman diagram does not modify the corresponding invariant.

To illustrate the use of the method exposed in this work we analyse
a simple observable, the CP-violating asymmetry in $Z\to \bar b s$ and
$Z\to b\bar s$ decays. Although the example is fully developed in
Appendix \ref{APP:Zbs}, we show in \fig{Fig:Zbs} (drawn using
\texttt{Jaxodraw} \cite{Binosi:2003yf}) one of the contributions to the asymmetry:
the interference between the standard amplitude of \fig{Zbs:fig:a} and
the new amplitude of \fig{Zbs:fig:b}, closing a {\em flavour path} and
joining both diagrams. The closed {\em flavour path} shadowed in
\fig{Fig:Zbs} gives the invariant trace (reading clockwise):
$\tr{\PdL{2} F(H_u) \PdL{3} G^\dagger(\HQ)}$; as explained in Appendix
\ref{APP:Zbs}, $F(H_u)$ and $G(\HQ)$ are loop functions and the
circles `$\circ$' over the $s$ and $b$ quark lines correspond to the
flavour projectors $\PdL{2}$ and $\PdL{3}$.

\begin{figure}[htb]
\begin{center}
\subfigure[SM amplitude\label{Zbs:fig:a}]{\raisebox{-44pt}{\epsfig{file=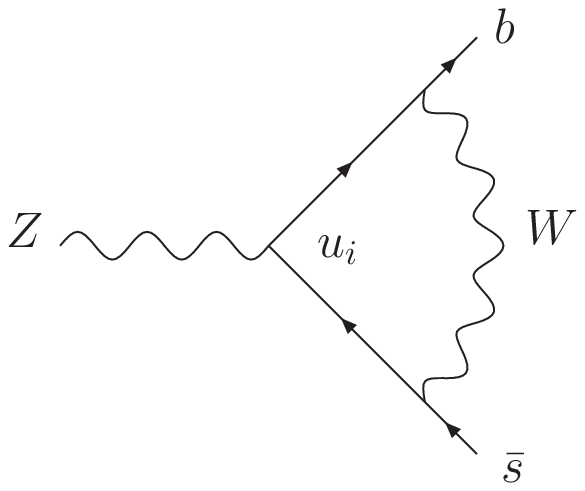,width=0.25\textwidth}}}$\quad\times\quad$\subfigure[MSSM amplitude\label{Zbs:fig:b}]{$\left(\raisebox{-45pt}{\epsfig{file=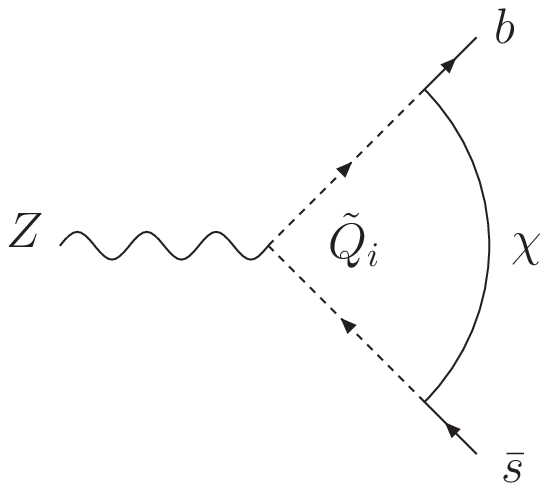,width=0.25\textwidth}}\right)^\ast$}
{\epsfig{file=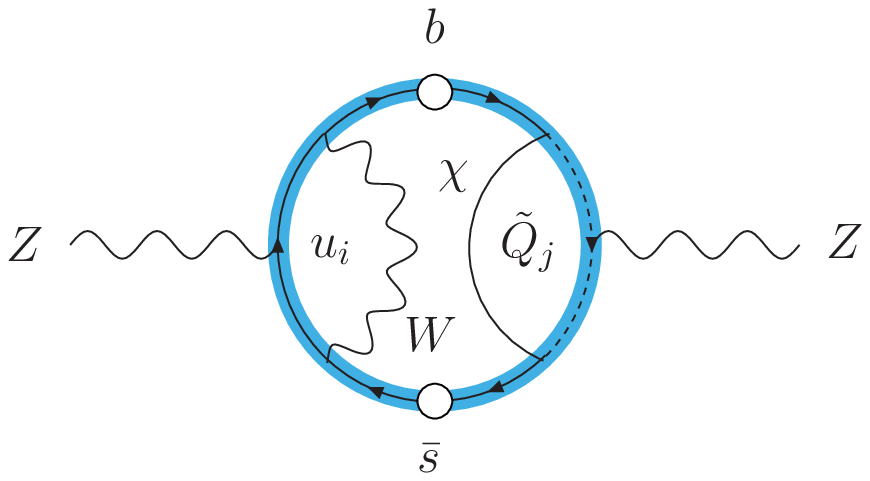,width=0.5\textwidth}}
\end{center}
\caption{Interference term contributing to $\Gamma(Z\to b\bar s)-\Gamma(Z\to\bar bs)$.}
\label{Fig:Zbs}
\end{figure} 

At this point we already have the necessary tools to choose our basis
of 4 independent invariants that fix all the observable phases in our
MSSM model and to relate them to physical observables. In the first place
we notice that we only have to consider invariants built with
projector operators. This is because any invariant,
including full flavour structures, $H_a$ with $a=u,d,\tilde Q$, can
always be written as a linear combination of invariants built with
projectors and masses\footnote{Strictly speaking, after electroweak symmetry breaking, left-handed squarks mix with right-handed squarks in a $6\times 6$ mass matrix and the chirality of the eigenvalues is not well defined. Here we can safely neglect the small left--right mixing proportional to Yukawa couplings. This issue will be further addressed in \cite{WorkInProgress}.}:

\begin{equation}
\label{fullvsproj}
\tr{(H_{u})^a (\HQ)^b (H_{d})^c \dots} = \sum_{\alpha\beta\gamma} (\masaU{\alpha})^a (\masaQ{\beta})^b (\masaD{\gamma})^c~ \tr{\PuL{\alpha} \PQ{\beta} \PdL{\gamma} \dots}.
\end{equation}
The first complex invariant we can build involves, at least, three different projectors, $\tr{\PuL{i} \PdL{j} \PQ{k}}$. However, it is more convenient to consider invariants with four matrices, for instance $\tr{\PuL{i} \PdL{j} \PQ{k} \PdL{l}}$. In fact, these four projector invariants correspond directly to the familiar rephasing invariant quartets of mixing matrices. Writing our hermitian matrices in terms of masses and relative misalignments,
\begin{equation}
H_d = D_d, \qquad H_u = V^\dagger D_u V, \qquad \HQ = U^\dagger D_{\tilde Q} U,
\end{equation}
where $D_a$ are diagonal matrices with eigenvalues $m^2_{a_i}$. We have 
\[
\tr{\PuL{i} \PdL{j} \PQ{k} \PdL{l}} = \V{ij} \U{kl} \Vc{il}\Uc{kj}.
\]
For this reason we will select our independent invariants from the invariants with four projectors. All other complex invariants can be written in terms of four matrices invariants using the
techniques in Appendix \ref{APP:traces}. In particular, using $\mathbf 1 = \sum_l \PdL{l}$, it is trivial to write invariants of three matrices in terms of four matrices invariants, $\tr{\PuL{i} \PdL{j} \PQ{k}} = \sum_l \tr{\PuL{i} \PdL{j} \PQ{k} \PdL{l}}$. 

The 4 projector invariants will involve at least two projectors of the same kind. They can be one of the following structures:
\begin{align} 
&\tr{\PuL{i} \PdL{j} \PuL{k} \PdL{l}},\quad\tr{\PuL{i} \PQ{j} \PuL{k} \PQ{l}},\quad\tr{\PQ{i} \PdL{j} \PQ{k} \PdL{l}}\label{set01}\\
&\tr{\PuL{i} \PdL{j} \PQ{k} \PdL{l}},\quad\tr{\PuL{i} \PQ{j} \PdL{k} \PQ{l}},\quad\tr{\PuL{i} \PdL{j} \PuL{k} \PQ{l}}.\label{set02}
\end{align} 
As shown in Appendix \ref{APP:traces}, all these different structures can be reduced to three families of invariants:
\begin{align}
\label{invfam}
\Jota{V}{ij}{kl} &\equiv \tr{\PuL{i} \PdL{j} \PuL{k} \PdL{l}} = \V{ij} \V{kl} \Vc{il} \Vc{kj} \nn \\ 
\Jota{U}{ij}{kl} &\equiv \tr{\PQ{i} \PdL{j} \PQ{k} \PdL{l}} = \U{ij} \U{kl} \Uc{il} \Uc{kj} \nn \\ 
\IInv{ij}{kl} &\equiv \tr{\PuL{i} \PdL{j} \PQ{k} \PdL{l}} =  \V{ij} \U{kl} \Vc{il} \Uc{kj}.
\end{align}
From here it is clear that $\Jota{V}{ij}{kl}$ are the familar
rephasing invariant quartets of the CKM mixing matrix (notice that $V$
corresponds to the CKM mixing matrix). It is well known that all the
quartets we can build have the same imaginary part and therefore we
only need one of them plus the moduli of the CKM elements to fix all
of them. The same is true for the quartets $\Jota{U}{ij}{kl}$,
although this time in terms of the relative misalignment between
squark doublets and down quarks. Therefore we choose as independent
quartets $\Jota{V}{32}{23}$ and $\Jota{U}{32}{23}$.

There are similar properties relating the different $\IInv{ij}{kl}$ quartets. Using the properties listed in Appendix \ref{APP:traces}, all these $\IInv{ij}{kl}$, $\Jota{V}{ij}{kl}$ and $\Jota{U}{ij}{kl}$ can be written in terms of only four independent quartets, which we choose to be
\begin{align}\label{independents}
\Jota{V}{32}{23} \equiv \tr{\PuL{3} \PdL{2} \PuL{2} \PdL{3}}, & \quad\Jota{U}{32}{23} \equiv \tr{\PQ{3} \PdL{2} \PQ{2} \PdL{3}},\nn\\
\IInv{33}{32}\equiv \tr{\PuL{3} \PdL{3} \PQ{3} \PdL{2}},& \quad\IInv{32}{31}\equiv \tr{\PuL{3} \PdL{2} \PQ{3} \PdL{1}},
\end{align}  
plus squared moduli of elements of the mixing matrices. These four
invariants constitute a basis of linearly independent complex
invariants. Any other complex invariant we can build in this theory
can be uniquely expressed as a linear combination of these four
invariants with coefficients proportional to masses and moduli of
elements of the mixing matrices. This is one of the key results of
this work as we are now able to relate unambiguously all the possible
CP-violating quantities of the theory and therefore make predictions
on different observables.

The last step is to relate these four independent invariants to
physical observables where they can be measured. So far supersymmetric
particles have not been directly observed and we will probably have to
wait until the LHC is in operation before we can analyse processes with SUSY particles
as external states. In the meantime, we can use FCNC and CP-violation
experiments to measure new contributions with SUSY particles running
in the loops. Consequently we choose our four independent observables
within this class of processes.

At the moment, CP violation has only been observed in neutral kaon and
neutral $B$ systems. These measurements seem to be consistent
with a Standard Model interpretation of the observed CP violation
\citeCKMfits. Nevertheless, any extension of the SM predicts some departure 
from the SM expectations once the experimental and theoretical precision is 
improved. On the other hand, the CKM Jarlskog quartet is also included in 
our independent set of
invariants and must be determined from the experimental
data. Therefore it is convenient to include in our set of observables
the two best experimental determinations of CP violation, indirect CP
violation in the neutral kaon system -- $\varepsilon_K$ -- and the CP
asymmetry in $B^0 \to J/\psi K_S$.

\begin{figure}[htb]
\begin{center}
\subfigure[SM amplitude\label{Kneutral:fig:a}]{\raisebox{-50pt}{\epsfig{file=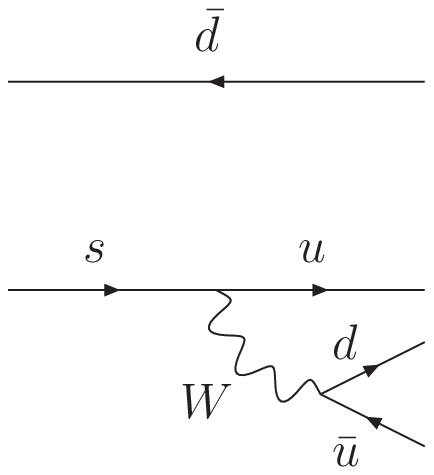,width=0.25\textwidth}}}$\quad\times\quad$\subfigure[MSSM in the $\bar K^0\to K^0$ box\label{Kneutral:fig:b}]{$\left(\raisebox{-53pt}{\epsfig{file=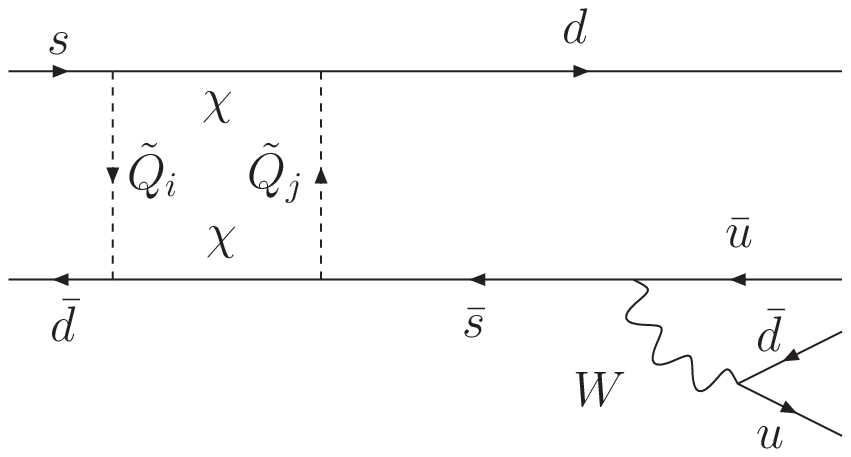,width=0.5\textwidth}}\right)^\ast$}
{\epsfig{file=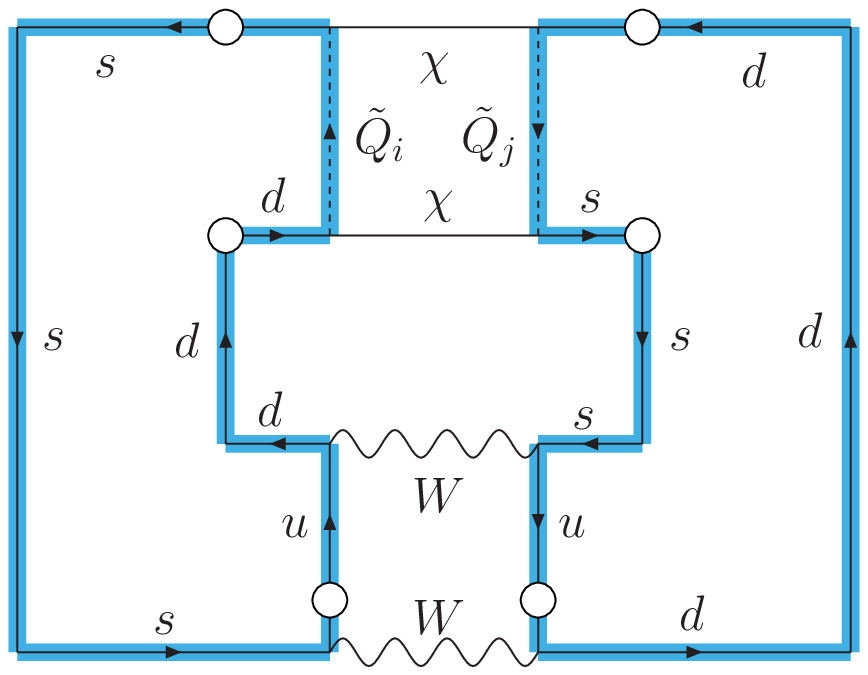,width=0.5\textwidth}}
\end{center}
\caption{CP violating contribution in $K^0\to\pi\pi$.}
\label{Fig:epsk}
\end{figure}

$\varepsilon_K$ corresponds to a particular combination of neutral
kaon decay rates: $K_L$ and $K_S$ decay rates with $I=0$, so that we
select CP violation in $K^0$--$\bar K^0$ mixing. One contribution to
these decays is, for example, the tree-level $K^0 \to \pi^+ \pi^-$ and
the mixing-mediated decay $K^0 \to \bar K^0 \to \pi^+ \pi^-$.  In this
case the external particles are two final-state pions and both the
$K^0$ and the $\bar K^0$, as we select explicitly CP violation in
$K^0$--$\bar K^0$ mixing. Therefore we need two $P^u_1$ and two
$P^d_1$ projectors corresponding to the pions, also two $P^d_2$
projectors and two $P^d_1$ projectors, the projectors corresponding to
the $K^0$ and $\bar K^0$. Naturally these processes will have
contributions from SM loops and new contributions from the virtual
SUSY particles 
\citeSUSYKMix. In \fig{Fig:epsk} we show one of the 
contributions to
the interference between the tree-level SM amplitude
(\fig{Kneutral:fig:a}) and the SUSY-mixing-mediated
(\fig{Kneutral:fig:b}) amplitude, that is, the leading contribution
beyond the SM. The invariant corresponding to this contribution is
easily obtained: starting from an $s$ quark projector, the
\emph{flavour path} (shadowed) in \fig{Fig:epsk} reads:
$\left[\tr{\PdL{2}\PuL{1}\PdL{1}\HQ}\right]^2$. More accurately the exact structure would be an arbitrary function $F(\HQ,\HQ)$, where $F(m_{Qi}^2,m_{Qj}^2)=\sum_{m,n} C_{mn}m_{Qi}^{2m},m_{Qj}^{2n}$ will be the corresponding Inami-Lim function \cite{Inami:1981fz}, see Appendix \ref{APP:Zbs} for an example. For simplicity, in the following we will only consider the leading term of these expansions.

Other contributions to the interference follow from \fig{Fig:epsk} by crossing internal lines in the $K^0$--$\bar K^0$ (box) mixing; they give rise to a second trace: $\tr{\PdL{2}\PuL{1}\PdL{1}\HQ\PdL{2}\PuL{1}\PdL{1}\HQ}$. As in Appendix \ref{APP:traces}, we easily reduce
\[
\tr{\PdL{2}\PuL{1}\PdL{1}\HQ\PdL{2}\PuL{1}\PdL{1}\HQ}=\left[\tr{\PdL{2}\PuL{1}\PdL{1}\HQ}\right]^2.
\]
Thus all the interference terms with SUSY particles running in the neutral kaon mixing share a flavour invariant structure. SM contributions involve up quarks and $W$'s in the mixing, thus giving terms proportional to:
\[
\left[\tr{\PdL{2}\PuL{1}\PdL{1}H_u}\right]^2.
\]
Consequently $\varepsilon_K$ depends on the following complex invariants:
\begin{equation}
\varepsilon_K=C_{\text{SM}}^{\varepsilon_K}\im{\tr{\PdL{2}\PuL{1}\PdL{1}H_u}^2}+C_{\text{MSSM}}^{\varepsilon_K}\im{\tr{\PdL{2}\PuL{1}\PdL{1}\HQ}^2},\label{EpsilonK:inv01}
\end{equation} 
where $C_{\text{SM}}^{\varepsilon_K}$ and $C_{\text{MSSM}}^{\varepsilon_K}$ are real coefficients that depend on coupling constants and real invariants.

The complex invariants relevant to the description of CP violation in
$B^0 \to J/\psi K_S$ are obtained as in the $\varepsilon_K$ case. The
only difference is the presence of an additional neutral meson mixing
because we have both $B^0$--$\bar B^0$ and $K^0$--$\bar K^0$ mixings,
implying that the analogue of \fig{Fig:epsk} involves an additional
box contribution. The $B^0 \to J/\psi K_S$ asymmetry is $A_{\text{CP}}(J/\psi
K_S)\propto \sin(2\varphi_{J/\psi K_S})$ where $\varphi_{J/\psi K_S}$
is given by
\begin{equation}
\varphi_{J/\psi K_S}\equiv \arg\left\{\sum_{i,j=u,\tilde Q}C_{ij}^{J/\psi K_S}\left(\tr{\PdL{1}H_i\PdL{3}\PuL{2}}\tr{\PdL{1}\PuL{2}\PdL{2}H_j}\right)^2\right\}\label{JPsiKs:inv01}
\end{equation}
and $C_{ij}^{J/\psi K_S}$ are, again, real coefficients that depend on
coupling constants and real invariants. In this case, we expect a
large contribution from the SM to this phase. However, a sizeable SUSY
contribution proportional to $\tr{\PdL{1}\HQ\PdL{3}\PuL{2}}$ is still
possible 
\citeSUSYBPsiK~and can play a relevant role in the unitarity 
triangle fit.

The third observable we are going to choose is the CP asymmetry in
$B_s \to J/\psi \Phi$ or $B_s \to D_s^+ D_s^-$. Notice that both
processes correspond exactly to the same decays at the quark level and
hence give rise to the same invariant.  This channel is especially
interesting for several reasons. First, many realizations
of supersymmetry can give a sizeable contribution to $B_s$--$\bar B_s$
mixing with a large phase 
\citeSUSYBsMix. Then, the SM 
contribution to
the CP asymmetry is very small and therefore a sizeable CP asymmetry
would be a signal of new physics.  Finally, this asymmetry is accessible
at $B$-physics experiments at hadron colliders such as LHCb or BTeV.

In this case, the diagrams are analogous to the $\varepsilon_K$
diagrams shown in \fig{Fig:epsk}. We also have a SM
contribution to the mixing and a new contribution from SUSY. The
corresponding invariants are $\im{\tr{\PdL{2}\PuL{2}\PdL{3}H_u}^2}$
for the SM contribution and $\im{\tr{\PdL{2}\PuL{2}\PdL{3}\HQ}^2}$ for
the MSSM contribution. This CP asymmetry is approximately dominated by
the tree-level decay amplitude \cite{Dunietz:2000cr}, and therefore
$A_{\text{CP}}(B_s \to D_s^+ D_s^-) \propto \sin (2 \varphi_{D_s^+ D_s^-})$.
The presence of a single mixing ($B^0$--$\bar B^0$) simplifies the analogue
of \eq{JPsiKs:inv01} and this phase is given by
\begin{equation}
\sin (\varphi_{D_s^+ D_s^-}) =\frac{\im{\tr{\PdL{2}\PuL{2}\PdL{3}H_u}^2}+C_{S}~\im{\tr{\PdL{2}\PuL{2}\PdL{3}\HQ}^2}}{\abs{\tr{\PdL{2}\PuL{2}\PdL{3}H_u}^2+C_{S}~\tr{\PdL{2}\PuL{2}\PdL{3}\HQ}^2}}.\label{DspDsm:inv01}
\end{equation}
The coefficient $C_S$ takes into account the differences in (real)
couplings and masses from the SM and the new SUSY contributions; it is known from other CP-conserving measurements. As said
above, the SM contribution to this asymmetry is small:
\begin{equation}
\varphi_{\text{SM}} = \arg\left\{\tr{\PdL{2}\PuL{2}\PdL{3}H_u}^2 \right\}\simeq {\mathcal{O}}(\lambda_c^2),
\end{equation}
with $\lambda_c$ the Cabibbo angle. Thus, in practice, this contribution can be safely neglected in the presence of a sizeable new physics contribution.

Finally, we need a fourth observable to obtain our four independent
invariants. The choice now is more difficult, and there is no clear
option. However, we choose the CP asymmetry in the $b \to s \gamma$
decay, which is already being measured at the $B$ factories
\citeBsgExp~and corresponds to a new invariant, independent
of the invariants involved in the previous observables 
\citeBsgTeo. Notice that
this process entails a change in the chirality of the down quarks,
i.e. it is a transition $b_R \to s_L \gamma$. This implies that we now 
need a right-handed projector $\PdR{3}$; however, using \eq{RightProj}, 
we have $\Yd\PdR{3}\Ydd = m_b^2 \PdL{3}$. Therefore, with the exception
of this additional quark mass, the diagrams involved are completely
analogous to the diagrams in the $Z \to b \bar s$ asymmetry and we have,
\begin{equation}
A_{\text{CP}}(b\to s \gamma) \propto \imtr{\PdL{3} H_u \PdL{2} H_{\tilde Q}}.\label{bsgamma:inv01}
\end{equation}

In summary, in Eqs. (\ref{EpsilonK:inv01})--(\ref{bsgamma:inv01}) we have four observables that can be expressed as functions of our four independent invariants using the relations of Appendix A.
Therefore we have four equations and four unknowns and we can fix
completely the four CP-violating invariants of our MSSM. This implies that any
other CP violation observable in this model is already fixed in terms
of our four invariants and masses or moduli of mixing angles. 

For instance, we can now calculate in our model the CP asymmetry in
the $B_d \to \phi K_S$ decay, which could show a discrepancy from the SM
expectations 
\citeSUSYBsMix. In this case the relevant invariant,
assuming that the large SUSY contribution is in the decay amplitude while the $B$--$\bar B$ mixing is SM-dominated,
would be
\begin{eqnarray}
\sin (\varphi_{\phi K_S}) &\propto& \imtr{\PdL{3} \HQ \PdL{2} H_u \PdL{1} H_u} \\
&\simeq& \frac{m_t^2}{|V_{tb}|^2}~ \imtr{\PdL{3}\HQ \PdL{2} \PuL{3}}
\re{\tr {\PdL{1} H_{u} \PdL{3} \PuL{3}}}. \nn
\end{eqnarray}
So, it is clear that this asymmetry in our model is directly related
to the CP asymmetry in $b \to s \gamma$ decays. Note that this is
only due to the fact that there are no other Left-Right 
couplings in our reduced MSSM model, apart from the usual Yukawas. Naturally, in a
complete MSSM, this relation may be destroyed by these additional
couplings. This kind of relations can be extended to any other CP-violating observable in the theory, for instance new SUSY contributions to
$\varepsilon^\prime/\varepsilon$ \cite{Khalil:2000ci}, 
$K\to \pi \nu \bar \nu$ 
\citeKPiNuNu, and possibly
CP asymmetries at future linear colliders 
\cite{Barger:2001nu,SUSYCollider}.

Let us now briefly discuss the realistic experimental determination of these 
observables.  First, we must emphasize that the situation regarding possible 
new physics contributions in B--factories has changed dramatically in recent 
times. Babar \cite{BABAR:gamma} and Belle \cite{BELLE:gamma} have presented for the
first time a measurement of the phase $\gamma =\arg \left(
-V_{ud}V_{ub}^{\ast }V_{cd}^{\ast }V_{cb}\right) $ from the "tree-level"
decays $B^{\pm }\rightarrow DK^{\pm },$ $B^{\pm }\rightarrow D^{\ast }K^{\pm
}\rightarrow \left( D\pi ^{0}\right) K^{\pm }$, where the two paths to $D^{0}
$ or $\bar{D}^{0}$ interfere in the common decay channel $\bar{D}^{0},D^{0}\rightarrow K_{S}\pi ^{+}\pi ^{-}$. This measurement corresponds
to the determination of the pure SM phase in B decays,
$\arg [\text{Tr}(P_{3}^{u_{L}}P_{2}^{d_{L}}P_{2}^{u_{L}}P_{3}^{d_{L}})]$. 
Even more important: B factories will  achieve a measurement of $\gamma$ with a 
precision of a few degrees in the near future \cite{gamma:future}. Using this observable 
together with the tree level observables $\abs{V_{us}},\abs{V_{ub}}$ and $\abs{V_{cb}}$ allows a high precision determination of 
the full $V_{CKM}$ ``independent'' of the presence of any new physics that 
respects $3 \times 3$ unitarity, as the MSSM analysed here. 
At this point, any other FCNC and/or CP violating observable could be 
devoted to the search of new phases as deviations from this tree level 
measurement. In this scenario, it would be enough to use $\gamma $ together 
with $\epsilon _{K},A_{CP}(b\rightarrow s\gamma )$ and 
$\varphi _{J/\psi K_{S}}$.
As is well known, at present the first two FCNC observables 
agree with the SM prediction and, in principle, can only accommodate
a relatively small MSSM contribution.
Recently \cite{Botella:2005fc,Bona:2005vz} a model independent analysis of $\varphi _{J/\psi K_{S}}$ has
shown that there are two solutions for this phase, one of them clearly 
outside the SM. Nevertheless the statistical significance of this second 
solution is smaller than the SM one \cite{Charles:2004jd}, and it is possible that this second
solution gets even less significant when more data become available \cite{Botella:2005fc,Bona:2005vz,Charles:2004jd}. 
Thus we prefer to wait for an updated analysis before we can consider 
this possibility a genuine new physics hint.
Now the experimental determination of the CP violating phase in 
$B_{s}^{0}$--$\bar{B}_{s}^{0}$,  $\varphi _{D_{s}^{+}D_{s}^{-}}$ is going
to be of paramount importance, specially taking into account its small value
in the SM. A clear signal of deviations from the SM in $\varphi
_{D_{s}^{+}D_{s}^{-}}$ would be a very welcome ingredient in our
program. Otherwise, % In other case, 
MSSM contributions to FCNC and CP violation processes
will be relatively small corrections to the SM predictions and high precision
measurements will be required.  

So, it goes without saying that the experimental determination of these 
observables is very challenging, requiring sustained hard work along the 
next years in different experiments, both direct production of SUSY particles
at colliders and indirect searches at FCNC and CP violation experiments.
In our exposition, all the CP-conserving quantities such as masses and
moduli of the different mixing angles are supposed to be known, accurately 
enough, in order to perform the analysis of the CP-violating quantities. 
In this scenario we can use direct measurements at high energy colliders, 
such as the LHC, the ILC, etc, and measurements
at FCNC experiments to extract the relevant phases.
Nevertheless in this framework our program can only be realized through 
a long and iterative process with a synergetic high energy-FCNC interplay, 
in which the first steps will not produce very precise results 
(see reference \cite{Barger:2001nu} for an example of a realistic analysis of flavour 
independent phases at colliders).

Finally, we would like to relate our expressions with weak basis
invariants to the usual computations in supersymmetric models, both
in the mass insertion formalism 
\citeMI~and in the exact mass eigenstate
formalism working with flavour changing vertices. The mass insertion formalism is just a series expansion on the small off-diagonal elements of the squark mass matrices, useful without performing a full diagonalization of them. In first place, we must take into account that,
as shown in Appendix \ref{APP:Zbs}, our invariants can contain an
arbitrary function of the internal hermitian mass matrices. These
arbitrary functions are, in the Feynman diagram computations, the usual
loop functions. Therefore, all we have to do to relate our invariant
formalism with the usual Feynman diagram calculations is to express
the internal hermitian matrices in terms of projectors, which give us
the mixing matrices entering in the process, and combine the mass
eigenstates of these matrices in the corresponding loop functions. The squark mass eigenstates are a mixture of left and right-handed squarks; nevertheless it is still possible to express $M_Q^2$ as a linear combination of the 6 squark masses and the 6$\times$6 squark mixing matrices \cite{WorkInProgress}. 
Naturally, gauge couplings do not enter in our invariants, but at
least we can identify the gauge couplings associated with gauginos and
$W$ bosons, as explained in our rules to build flavour diagrams given above. 
The
translation to the mass insertion formalism is also straightforward
from here. In this case, we do not express the internal squark mass
matrix in terms of projectors, and replace the mass eigenstates by a
universal squark mass in the loop functions. Then the hermitian squark
mass matrix in the invariant plays the role of a new off-diagonal
flavour coupling and the usual mass insertion corresponds directly to
$(\delta^x_L)_{ij}= (\UxLdag \HQ \UxL)_{ij}/m^2_{\tilde Q}$. Notice that, 
as pointed out in \cite{Lebedev:2002wq}, the full invariant must contain
additional mixing matrices to be completely weak basis invariant.

%%%%%%%%%%%%%%%%%%%%%%%%%%%%%%%%%%%%%%%%%%%%%%%%%%
\section{Conclusions}\label{SEC:conclusions}
%%%%%%%%%%%%%%%%%%%%%%%%%%%%%%%%%%%%%%%%%%%%%%%%%%

In this work we have presented the complete machinery necessary to
find all the independent WBIs in any extension of the
Standard Model and to relate them to physical observables. We have
defined weak basis and rephasing invariance, and shown how any flavour
process in the Standard Model, and in particular any CP-violating
process, can be easily expressed in terms of WBIs. We
have introduced a graphical representation of these 
WBIs as a simple extension of the usual Feynman diagrams. As a
practical application, we have found all the independent observables
in a reduced version of the MSSM with only three flavour matrices. In
this model, we have been able to define a basis of four complex
invariants spanning all the observable phases in the model. Then we
have chosen four different physical processes to fix these
four invariants completely; from there, assuming we know the sparticles masses
and moduli of the mixings, we are able to make predictions on any
other CP-violating observable in the model.
 
This formalism can be applied to more complete models, as for
instance the full MSSM or any other extension of the SM with new
flavour structures.  This analysis will be presented in a future work
\cite{WorkInProgress}.

\section*{Acknowledgements}
This work has been supported by MEC
under FPA2002-00612. M.N.
acknowledges MEC for a fellowship and the warm hospitality
during his stays at Oxford and CERN, where parts of this work were done. The authors thank A. Santamaria and T. Hurth for useful discussions.

\appendix
\newpage
%%%%%%%%%%%%%%%%%%%%%%%%%%%%%%%%%%%%%%%%%%%%%%%%%%
\section{Traces and mixings}\label{APP:traces}
%%%%%%%%%%%%%%%%%%%%%%%%%%%%%%%%%%%%%%%%%%%%%%%%%%
In this appendix we prove that any complex invariant in our reduced
MSSM can be written as a linear combination of only four independent
complex invariants. Using \eq{fullvsproj} we can concentrate on invariants 
built only with projectors.

It is obvious that invariants with two projectors are always real and, in fact,
\eq{OBS:00}, they simply carry the moduli of elements of mixing matrices. 
We diagonalize the hermitian matrices, $H_u= \UuLdag 
\text{Diag}(m_{u_i}^2) \UuL$, $H_d= \UdLdag \text{Diag}(m_{d_i}^2) \UdL$ and 
$\HQ=\UQ\text{Diag}(m_{Q_i}^2)\UQdag $ and define the mixing matrices
$V\equiv\UuLdag~\UdL$ (just the CKM matrix) and $U\equiv\UQdag\UdL$. Then,
we obtain the invariant moduli
\begin{equation}
\tr{\PuL{i}\PdL{j}}=\abs{\V{ij}}^2\quad ;\quad
\tr{\PQ{k}\PdL{j}}=\abs{\U{kj}}^2.\label{MSSM:2tr:moduli}
\end{equation}

The next step is to consider invariants involving three different 
projectors, $\tr{\PuL{i}\PdL{j}\PQ{k}}$, which can have non-zero 
imaginary parts. Nevertheless, as $\mathbf{1}=\sum \text{Projectors}$, 
we can write
\begin{equation}
\tr{\PuL{i}\PdL{j}\PQ{k}}=\sum_\ell\tr{\PuL{i}\PdL{j}\PQ{k}\PdL{\ell}}\label{MSSM:3P:red}
\end{equation}
Any trace over 3 projectors can be expressed in terms of a sum of
traces over 4 projectors where one kind of projector appears
twice. 

Now, we have two kinds of invariants with 4 projectors. First, invariants
that involve the three sorts of projectors $\PdL{i},\PuL{i},\PQ{i}$:
\begin{equation}
\tr{\PuL{i}\PdL{j}\PQ{k}\PdL{\ell}},\qquad \tr{\PuL{i}\PdL{j}\PuL{k}\PQ{\ell}},\qquad \tr{\PuL{i}\PQ{j}\PdL{k}\PQ{\ell}},\label{MSSM:4P:eq02}
\end{equation}
then, invariants that involve only two sorts of projectors:
\begin{equation}
\tr{\PuL{i_1}\PdL{j_1}\PuL{i_2}\PdL{j_2}},\qquad \tr{\PdL{i_1}\PQ{j_1}\PdL{i_2}\PQ{j_2}},\qquad \tr{\PuL{i_1}\PQ{j_1}\PuL{i_2}\PQ{j_2}}.\label{MSSM:4P:eq01}
\end{equation}
As in the SM case,  using unitarity of the mixing matrix, each family of 
invariants in \eq{MSSM:4P:eq01} provides a single imaginary part.

Using the mixing matrices defined above, we define,
\begin{align}
\Jota{V}{i_1j_1}{i_2j_2}&\equiv\tr{\PuL{i_1}\PdL{j_1}\PuL{i_2}\PdL{j_2}}=\V{i_1j_1}\Vc{i_2j_1}\V{i_2j_2}\Vc{i_1j_2}\label{MSSM:4tr:quartets01}\\
\Jota{U}{k_1j_1}{k_2j_2}&\equiv\tr{\PQ{k_1}\PdL{j_1}\PQ{k_2}\PdL{j_2}}=\U{k_1j_1}\Uc{k_2j_1}\U{k_2j_2}\Uc{k_1j_2}\label{MSSM:4tr:quartets02}\\
\IInv{ij}{k\ell}&\equiv\tr{\PuL{i}\PdL{j}\PQ{k}\PdL{\ell}}=\V{ij}\Vc{i\ell}\U{k\ell}\Uc{kj}.
\label{MSSM:4tr:quartets03}
\end{align}

And now, with $\mathbf{1}=\sum \text{Projectors}$ and $\tr{A P_i B
P_i}=\tr{A P_i}\tr{B P_i}$ (for any projector $P_i$), we have:
\begin{eqnarray}
\tr{\PuL{i_1}\PdL{j_1}\PuL{i_2}\PQ{k}}&=& \sum_{j_2}
\tr{\PuL{i_1}\PdL{j_1}\PuL{i_2}\PQ{k}\PdL{j_2}}\qquad \qquad\nn \\& =&
\sum_{j_2}\Frac{\tr{\PuL{i_1}\PdL{j_1}\PuL{i_2}\PQ{k}\PdL{j_2}}
\tr{\PuL{i_2}\PdL{j_2}}}{\tr{\PuL{i_2}\PdL{j_2}}} \nn\\&=&
\sum_{j_2}\Frac{\tr{\PuL{i_2}\PQ{k}\PdL{j_2}}\tr{\PuL{i_2}\PdL{j_2}\PuL{i_1}
\PdL{j_1}}}{\tr{\PuL{i_2}\PdL{j_2}}} \nn \\ &=& \sum_{j_2,j_3}
\Frac{\tr{\PuL{i_2}\PdL{j_3}\PQ{k}\PdL{j_2}}\tr{\PuL{i_2}\PdL{j_2}\PuL{i_1}
\PdL{j_1}}}{\tr{\PuL{i_2}\PdL{j_2}}}\nn \\& =& \sum_{j_2,j_3}
\Frac{\IInv{i_2 j_3}{k j_2}~ \Jota{V}{i_2 j_2}{i_1 j_1}}{\abs{\V{i_2
j_2}}^2}. \label{MSSM:4P:red01}
\end{eqnarray}
In a similar way, we obtain
\begin{equation}
\tr{\PuL{i}\PQ{k_1}\PdL{j_1}\PQ{k_2}}=\sum_{j_2,j_3}\frac{\IInv{i j_2}{k_1 j_3} ~\Jota{U}{k_1 j_1}{k_1 j_3}}{\abs{\U{k_1 j_3}}^2} \label{MSSM:4P:red02}
\end{equation}
\begin{equation}
\tr{\PQ{k_1}\PuL{i_1}\PQ{k_2}\PuL{i_2}}=\sum_{j_1,j_2,j_3,j_4}
\Frac{\IInv{i_1 j_1}{k_1 j_2}~\IInv{i_2 j_1}{k_2 j_3}~ \IInv{i_2
j_4}{k_1 j_1}}{\abs{\U{k_1 j_1}}^2\abs{\V{i_2 j_1}}^2}.
 \label{MSSM:4P:red03}
\end{equation}
Using these relations, the invariant traces in Eqs. (\ref{MSSM:4P:eq01}) and 
(\ref{MSSM:4P:eq02}) can be reduced to three different families,
$\tr{\PuL{i_1}\PdL{j_1}\PuL{i_2}\PdL{j_2}}$,
$\tr{\PdL{i_1}\PQ{j_1}\PdL{i_2}\PQ{j_2}}$ (each one providing a single
imaginary part) and $\tr{\PuL{i}\PdL{j}\PQ{k}\PdL{\ell}}$.
Higher-order invariants (with more than 4 projectors) are easily reduced to 
the ones considered through the same method. 

With $\tr{\PuL{i_1}\PdL{j_1}\PuL{i_2}\PdL{j_2}}$ and
$\tr{\PdL{i_1}\PQ{j_1}\PdL{i_2}\PQ{j_2}}$ providing two independent
imaginary parts, we turn our attention to
$\tr{\PuL{i}\PdL{j}\PQ{k}\PdL{\ell}}$, particularly to the number of
independent imaginary parts in this family of invariants. The
interesting relations now are:
\begin{equation}
\tr{\PuL{{\mathbf i}}\PdL{j}\PQ{k}\PdL{\ell}} = \Frac{\tr{\PuL{i}\PdL{j}\PuL{m}\PdL{\ell}} \tr{\PuL{{\mathbf m}}\PdL{j}\PQ{k}\PdL{\ell}}}{\tr{\PuL{m}\PdL{\ell}}\tr{\PuL{m}\PdL{j}}} \label{MSSM:4P:red04}
\end{equation}
\begin{equation}
\tr{\PuL{i}\PdL{j}\PQ{{\mathbf k}}\PdL{\ell}} = \Frac{\tr{\PQ{k}\PdL{\ell}\PQ{m}\PdL{j}} \tr{\PuL{i}\PdL{j}\PQ{{\mathbf m}}\PdL{\ell}}}{\tr{\PQ{m}\PdL{\ell}}\tr{\PQ{m}\PdL{j}}}  \label{MSSM:4P:red05}
\end{equation}
\begin{equation}
\tr{\PuL{i}\PdL{{\mathbf j}}\PQ{k}\PdL{\boldsymbol{\ell}}} = \Frac{\tr{\PuL{i}\PdL{{\mathbf j}}\PQ{k}\PdL{{\mathbf a}}} \tr{\PuL{i}\PdL{{\mathbf a}}\PQ{k}\PdL{{\boldsymbol{\ell}}}}}{\tr{\PuL{i}\PdL{a}}\tr{\PQ{k}\PdL{a}}}. \label{MSSM:4P:red06}
\end{equation}
Equation (\ref{MSSM:4P:red04}) allows us to select an arbitrary $i$ in
$\tr{\PuL{i}\PdL{j}\PQ{k}\PdL{\ell}}$; the second relation,
\eq{MSSM:4P:red05}, allows us to select an arbitrary $k$ in
$\tr{\PuL{i}\PdL{j}\PQ{k}\PdL{\ell}}$. As the exchange
$j\leftrightarrows \ell$ amounts to a conjugation, that is
$\tr{\PuL{i}\PdL{j}\PQ{k}\PdL{\ell}}=$
$\left[\tr{\PuL{i}\PdL{\ell}\PQ{k}\PdL{j}}\right]^\ast$, only 3
different $\tr{\PuL{i}\PdL{j}\PQ{k}\PdL{\ell}}$ are independent, the
ones in which $j\neq \ell$. This number is further reduced to 2 by the
third relation, \eq{MSSM:4P:red06}; these two invariants, together
with $\tr{\PuL{i_1}\PdL{j_1}\PuL{i_2}\PdL{j_2}}$ and
$\tr{\PdL{i_1}\PQ{j_1}\PdL{i_2}\PQ{j_2}}$, span the 4 observable CP-violating phases that appear in this restricted MSSM.

\newpage
%%%%%%%%%%%%%%%%%%%%%%%%%%%%%%%%%%%%%%%%%%%%%%%%%%
\section{$Z\to b\bar s$ example}\label{APP:Zbs}
%%%%%%%%%%%%%%%%%%%%%%%%%%%%%%%%%%%%%%%%%%%%%%%%%%
As a simple illustrating example we will analyse the CP-violating rate asymmetry $\Gamma(Z\to b\bar s)-\Gamma(Z\to \bar bs)$; let us consider the simplest complex invariant traces that may appear in this observable. As we fix $b$ and $s$ external quarks this requires the presence of $\PdL{3}$ and $\PdL{2}$; there are no other projectors. With two down quark projectors the trace requires two additional matrices to exhibit an imaginary part; as the available matrices are $H_u$ and $M_Q^2$, we can expect the presence of the structures
\begin{align}
\tr{\PdL{2}f_1(H_u) \PdL{3}f_2(H_u)}\quad;\quad \tr{\PdL{2}g_1(M_Q^2) \PdL{3}g_2(M_Q^2)}\nn\\ \tr{\PdL{2}f_3(H_u)\PdL{3}g_3(M_Q^2)}\quad;\quad \tr{\PdL{2}g_4(M_Q^2)\PdL{3}f_4(H_u)},\label{Zbs:inv00}
\end{align}
where $f_i(H_u)$ and $g_i(M_Q^2)$ are functions of $H_u$ and $M_Q^2$ (loop functions). 

Let us consider the leading amplitude $A\equiv A(Z\to b\bar s)$:
\begin{equation}
A=~\raisebox{-45pt}{\epsfig{file=./Figs/Zbs_Amplitude_SM.eps,width=0.25\textwidth}}~~+~~\raisebox{-45pt}{\epsfig{file=./Figs/Zbs_Amplitude_MSSM.eps,width=0.25\textwidth}}\label{Zbs:amplitude01}
\end{equation}
The first kind of contribution is the SM one; the second one is the first SUSY contribution in our simple MSSM, with squarks and gauginos running in the loop. Schematically:
\begin{multline}
\abs{A}^2=~\raisebox{-40pt}{\epsfig{file=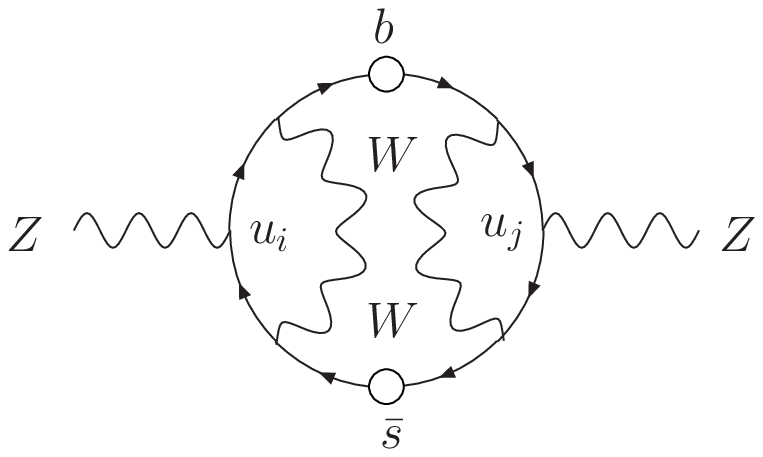,width=0.35\textwidth}}~~+~~\raisebox{-40pt}{\epsfig{file=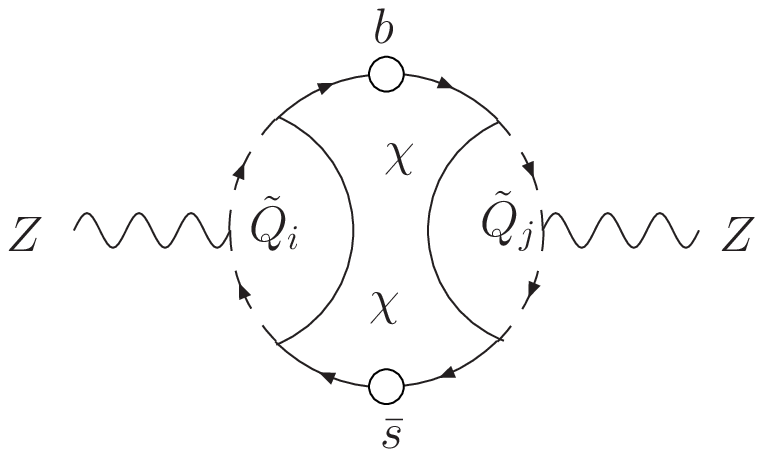,width=0.35\textwidth}}~~+\\ ~~ 2\text{Re}\left[\raisebox{-40pt}{\epsfig{file=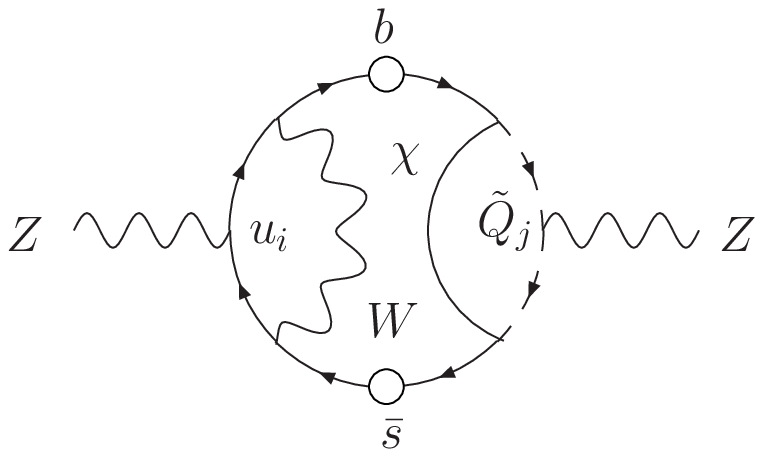,width=0.35\textwidth}}\right]\label{Zbs:amplitude02}
\end{multline}
Notice that the insertion of '$\circ$' in the diagrams recalls the fact that the $b$ and $\bar s$ are external states. In terms of invariant traces,
\begin{equation}
\raisebox{-40pt}{\epsfig{file=./Figs/Zbs_sAmplitude_SM_SM.eps,width=0.35\textwidth}}~~\to~~\raisebox{-33pt}{\epsfig{file=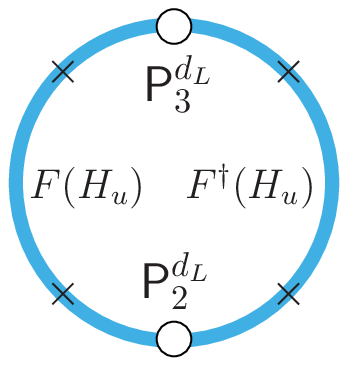,width=0.18\textwidth}}~~=\tr{\PdL{2}F(H_u)\PdL{3}F^\dagger(H_u)}\label{Zbs:trace01}
\end{equation}
\begin{equation}
~\raisebox{-40pt}{\epsfig{file=./Figs/Zbs_sAmplitude_MSSM_MSSM.eps,width=0.35\textwidth}}~~\to~~\raisebox{-33pt}{\epsfig{file=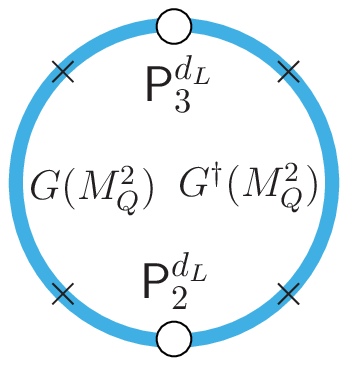,width=0.18\textwidth}}~~=\tr{\PdL{2}G(M_Q^2)\PdL{3}G^\dagger(M_Q^2)}\label{Zbs:trace02}
\end{equation}
\begin{equation}
\raisebox{-40pt}{\epsfig{file=./Figs/Zbs_sAmplitude_SM_MSSM.eps,width=0.35\textwidth}}~~\to~~\raisebox{-33pt}{\epsfig{file=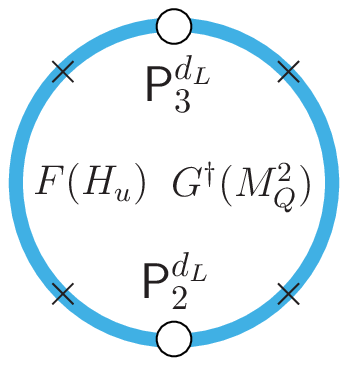,width=0.18\textwidth}}~~=\tr{\PdL{2}F(H_u)\PdL{3}G^\dagger(M_Q^2)}\label{Zbs:trace03}
\end{equation}
That is
\begin{multline}
\abs{A}^2=\tr{\PdL{2}F(H_u)\PdL{3}F^\dagger(H_u)}+\tr{\PdL{2}G(M_Q^2)\PdL{3}G^\dagger(M_Q^2)}+\\ 2\text{Re}\left[\tr{\PdL{2}F(H_u)\PdL{3}G^\dagger(M_Q^2)}\right]\label{Zbs:amplitude03}
\end{multline}
Similarly the amplitude $\bar A=A(Z\to \bar bs)$ is
\begin{multline}
\abs{\bar A}^2=\tr{\PdL{2}F^\dagger(H_u)\PdL{3}F(H_u)}+\tr{\PdL{2}G^\dagger(M_Q^2)\PdL{3}G(M_Q^2)}+\\ 2\text{Re}\left[\tr{\PdL{2}F^\dagger(H_u)\PdL{3}G(M_Q^2)}\right].\label{Zbs:amplitude04}
\end{multline}
Decomposing the loop functions in dispersive and absorptive pieces,
\begin{equation}
F(H_u)=F_{Dis}(H_u) +i F_{Abs}(H_u)\quad ; \quad G(M_Q^2)=G_{Dis}(M_Q^2) + i G_{Abs}(M_Q^2),\label{Zbs:functions}
\end{equation}
we can simplify the CP asymmetry $A_{\text{CP}}=\abs{A}^2-\abs{\bar A}^2$:
\begin{align}
& \abs{A}^2-\abs{\bar A}^2= \nn\\ &4\im{\tr{\PdL{2}F_{Dis}(H_u)\PdL{3}F_{Abs}(H_u)}}+4\im{\tr{\PdL{2}G_{Dis}(M_Q^2)\PdL{3}G_{Abs}(M_Q^2)}}+\nn\\
&4\im{\tr{\PdL{2}F_{Dis}(H_u)\PdL{3}G_{Abs}(M_Q^2)}}-4\im{\tr{\PdL{2}F_{Abs}(H_u)\PdL{3}G_{Dis}(M_Q^2)}}.\label{Zbs:asymmetry00}
\end{align}
By expanding the different functions, for example $F_{Dis}(H_u)=\sum_j F_{Dis}(m_{u_j}^2)\PuL{j}$, we can write
\begin{align}
&\abs{A}^2-\abs{\bar A}^2=\nn\\ & 4\sum_{i,j}F_{Dis}(m_{u_i}^2)F_{Abs}(m_{u_j}^2)\im{\tr{\PdL{2}\PuL{i}\PdL{3}\PuL{j}}}+\nn\\
& 4\sum_{i,j}G_{Dis}(m_{Q_i}^2)G_{Abs}(m_{Q_j}^2)\im{\tr{\PdL{2}\PQ{i}\PdL{3}\PQ{j}}}+\nn\\
& 4\sum_{i,j}\left[F_{Dis}(m_{u_i}^2)G_{Abs}(m_{Q_j}^2)-F_{Abs}(m_{u_i}^2)G_{Dis}(m_{Q_j}^2)\right]\im{\tr{\PdL{2}\PuL{i}\PdL{3}\PQ{j}}}.\label{Zbs:asymmetry01}
\end{align}
The asymmetry is thus easily written in terms of different irreducibly complex invariants; \eq{Zbs:asymmetry01} is further reduced as there are no absorptive parts in the loops containing squarks or top quarks: $F_{Abs}(m_{t}^2)=G_{Abs}(m_{Q_i}^2)=0$.

\end{document}